\documentclass[draft, journal,onecolumn,12pt]{IEEEtran}
\usepackage{amssymb,amsmath,epsfig,graphicx,theorem}
\usepackage{amsfonts}
\usepackage{bm}
\newtheorem{Theo}{Theorem}
\newtheorem{Lem}{Lemma}

\newtheorem{Def}{Definition}
\usepackage{epsfig,rotating,setspace,latexsym,amsmath,epsf,amssymb,bm}
\usepackage{cite}

\IEEEoverridecommandlockouts

\title{Deception with Side Information in Biometric Authentication Systems}
\author{
\IEEEauthorblockN{Wei Kang\qquad Daming Cao\qquad Nan Liu}\\
  \IEEEauthorblockA{School of Information Science and Engineering\\
Southeast University\\
Nanjing, China 210096\\
Email: wkang@seu.edu.cn, dmcao@seu.edu.cn, nanliu@seu.edu.cn}\\
\thanks{This paper was presented in part at IEEE International Symposium on Information Theory (ISIT), 2014.  This work is partially
supported by the National Basic Research Program of China
(973 Program 2012CB316004), the National Natural Science Foundation of China
under Grants $61271208$, $61201170$  and $61221002$,
the Research Fund of National Mobile Communications Research Laboratory,
Southeast University (No. 2014A02),
the Project-sponsored by SRF for ROCS, SEM and Qing Lan Project.}}

\begin{document}
\maketitle

\begin{abstract}
In this paper, we study the probability of successful deception of an uncompressed biometric authentication system with side information at the adversary. It represents the scenario where the adversary may have correlated side information, e.g.,~a partial finger print or a DNA sequence of a relative of the legitimate user. We find the optimal exponent of the deception probability by proving both the achievability and the converse. Our proofs are based on the connection between the problem of deception with side information and the rate distortion problem with side information at both the encoder and decoder.
\end{abstract}
\textbf{Keywords}:
Information theoretical security, Biometric authentication, Side information, Permutation, Blowing up lemma

\newpage
\section{Introduction}
The biometric authentication problem has been studied extensively in recent years. In a biometric authentication system, a biometric feature, e.g.,~a finger print, a DNA sequence, etc., of a legitimate user is measured and the measurement, called enrollment, is stored in a database. Later this biometric feature of the same user is measured and compared with the enrollment for authentication. Due to the randomness in the process, different measurements of the same biometric features of the same person can not be exactly the same. Thus, the authentication system needs to tolerate a certain level of distortion between the measurements taken in the enrollment stage and the authentication stage.

In a biometric authentication system, successful deception happens when an adversary impersonates a legitimate user by faking a biometric feature close enough to the enrollment and then deceives the authentication system. 
The first study on the deception probability in the biometric authentication system is \cite{Ahlswede:1997}, where the authors studied the deception in an authentication system where the enrollment is compressed. The authors obtained the optimal trade-off between the compression rate and the exponent of the probability of successful deception when the adversary has no side information. In the case with correlated side information at the adversary, achievability and converse results on the optimal trade-off were proposed in the paper,  however, they do not meet. 
A similar result was obtained in a recent paper \cite{Willems:2012}, where the optimal exponent of the deception probability has been given in both cases of uncompressed and compressed enrollment with no side information at the adversary.

A different direction in studying the performance of a biometric authentication system is to study the maximum number of legitimate users, called capacity, allowed in a biometric authentication system under a given tolerated distortion level. In \cite{Willems:2003}, the capacity is obtained if the enrollment is not compressed. Later, the capacity result is generalized to the case  where the enrollment is compressed \cite{Tuncel:2009}, and the trade-off between the compression rate and the capacity of the authentication system was studied.  The threat of a deception from an adversary was not considered in this line of work.

In this paper, we study the optimal exponent of the probability of successful deception of a biometric authentication system with uncompressed enrollment and side information at the adversary. It represents the scenario where the adversary may have correlated side information, e.g.,~a partial finger print of the legitimate user or a DNA sequence of a relative of the legitimate user. We provide the optimal exponent of the deception probability by providing the proofs of both the achievability and the converse. Our proofs are based on a connection between the problem of deception with side information and the rate distortion problem with side information at both the encoder and decoder.

The reminder of the paper is as follows. In the next section, we state the problem formulation and the main result. The proofs of the achievability and the converse are given in Section III and IV, respectively. Finally, a conclusion is given in Section V.

\section{Problem Formulation and Main Result}
\subsection{Problem Formulation}
Consider a pair of independent and identically distributed (i.i.d.) random sequences $X^n$ and $Y^n$, generated according to a joint distribution $P$, which is defined on a finite space $\mathcal{X}\times\mathcal{Y}$. Th random sequence $X^n$ represents the biometric enrollment in the system and $Y^n$ represents the side information at the adversary. We define a reconstruction space $\hat{\mathcal{X}}$ and a distortion function $d:\mathcal{X}\times\hat{\mathcal{X}}\mapsto \mathbb{R}^+\cup\{0\}$. The distortion between the sequences $x^n\in\mathcal{X}^n$ and $\hat{x}^n\in\hat{\mathcal{X}}^n$  is defined as 
\begin{align}
d(x^n,\hat{x}^n)\triangleq\frac{1}{n}\sum_{i=1}^nd(x_i,\hat{x}_i). \label{seqdis}
\end{align}
The legitimate user is successfully identified if the distortion between the measurements in the enrollment stage and the authentication stage does not exceed a certain level, say $\Delta$. We note that the probability of false rejection in our model is the same as the model without side information at the adversary. Therefore, we refer the readers to \cite{Ahlswede:1997,Willems:2012} for the derivation of the maximal probability of false rejection.

The adversary observes the side information $Y^n$ and tries to impersonate the legitimate user using a deception function 
$
f:\mathcal{Y}^n\mapsto\hat{\mathcal{X}}^n
$. 
We define the achievable deception exponent as follows.
\begin{Def}
A deception exponent $E$ is achievable under the distortion constraint $\Delta$ if there exists a deception function $f$ such that
\begin{align}
-\frac{1}{n}\log\mathsf{Pr}(d(X^n,f(Y^n))\le \Delta+\delta)\le E+\delta.\label{defi}
\end{align}
\end{Def}

In this paper, we are interested in the minimal achievable deception exponent, which is the best the adversary can do. Based on the minimal achievable deception exponent, the designer of the biometric authentication system can choose an appropriate $\Delta$ value that on one hand, limits the probability of successful deception below the tolerance level, and on the other hand, does not cause too large a probability of false rejection when the legitimate user is authenticated \cite{Ahlswede:1997, Willems:2012}.

\subsection{Rate-distortion with Side Information at Both the Encoder and Decoder}\label{WZ}
It turns out that finding the minimal achievable deception exponent with side information at the adversary is intimately related to the rate distortion problem with side information at both the encoder and decoder \cite{Wyner:1976}. Thus, in this subsection, we review the result for the rate distortion problem. 

Assume a pair of i.i.d sequences $X^n$ and $Y^n$ generated according to a joint distribution $Q$ defined on $\mathcal{X}\times\mathcal{Y}$, where $Q$ is not necessarily equal to $P$, which is defined in the previous subsection. The random sequence $X^n$ is the source sequence to be reconstructed at the decoder under a certain distortion constraint and $Y^n$ represents the side information available at both the encoder and decoder. The encoding function at the encoder is defined as
$
g:\mathcal{X}^n\times\mathcal{Y}^n\mapsto\{1,2,\dots,M\}, 
$
and the decoding function at the decoder is defined as
$
\varphi:\{1,2,\dots,M\}\times\mathcal{Y}^n\mapsto \hat{\mathcal{X}}^n
$. 
We denote the minimal achievable rate under distortion constraint $\Delta$ in the rate distortion problem with side information at both the encoder and decoder as $R_{SI}(Q,\Delta)$. From \cite{Wyner:1976}, we have
\begin{align}
R_{SI}(Q,\Delta) = \min_{V(\hat{x}|x,y): \mathsf{E}d(X,\hat{X})\le \Delta} I(X;\hat{X}|Y). \label{WZrate}
\end{align}

\emph{Remark}: The above rate distortion problem with side information at both the encoder and decoder can also be viewed as a special case of the Wyner-Ziv problem, i.e.,~the rate distortion problem with side information only at the decoder, as follows: in the Wyner-Ziv problem, view $(X^n,Y^n)$ jointly as the source sequence available at the encoder, view $Y^n$ as the side information at the decoder, and take the the distortion function in the Wyner-ziv problem as
$d(x,\hat{x})$, which is defined in the previous subsection, i.e., the distortion of $Y^n$ does not matter. 
By viewing the rate distortion problem with side information at both the encoder and decoder as a special case of the Wyner-Ziv problem, we can invoke the results of the  Wyner-Ziv problem, e.g.,~\cite[Theorem 16.5]{Csiszar:2011}, in later development. 

\subsection{Main Result}
The main result of this paper is the following theorem.
\begin{Theo} \label{main}
The deception exponent $E$ is achievable under the distortion constraint $\Delta$ if and only if 
\begin{align}
E\ge \min_{Q}\{D(Q||P)+R_{SI}(Q,\Delta)\}, \label{main_eq}
\end{align}
where $R_{SI}(Q,\Delta)$ is given in (\ref{WZrate}), the distribution $Q$ is defined on $\mathcal{X}\times\mathcal{Y}$ and $D(Q||P)$ represents the Kullback-Leibler divergence between the distributions $Q$ and $P$ \cite{Csiszar:2011}.
\end{Theo}
In the next two sections, we will show the proofs of the achievability and the converse of Theorem \ref{main} via the connection between the problem of deception with side information and the rate distortion problem with side information at both the encoder and decoder.  
%

\section{The Achievability}
In this section, we will show that there exists a deception function $f$ that can achieve the the deception exponent $D(Q||P)+R_{SI}(Q,\Delta)$ for the distortion constraint $\Delta$ and the distribution $Q$ defined on $\mathcal{X}\times\mathcal{Y}$. We will construct the deception function $f$ from the rate distortion code with side information at both the encoder and decoder.  


First, consider the rate distortion problem with side information at both the encoder and decoder as defined in subsection \ref{WZ}, where $(X^n, Y^n)$ are generated i.i.d. according to the distribution $Q$. Theorem 16.5 in \cite{Csiszar:2011} shows that for any sufficiently large $n$, and $0<\tau<1$, there exists a length $n$ code that achieves the rate $R_{SI}(Q,\Delta)$ and satisfies the distortion constraint with probability larger than $1-\tau$. More specifically, there exists a function pair $(g,\varphi)$ such that
\begin{align}
\frac{1}{n}\log||g||&\le R_{SI}(Q,\Delta)+\delta\label{WZr}\\
\mathsf{Pr}\left(d\left(X^n, \varphi(g(X^n,Y^n),Y^n)\right)\le\Delta+\delta\right)&\ge1-\tau\label{tau}
\end{align}

Define $A\subset\mathcal{X}^n\times\mathcal{Y}^n$ as the set of sequences $(x^n,y^n)$ that satisfies the distortion constraint $\Delta+\delta$ under $(g,\varphi)$, i.e.,
\begin{align}
A\triangleq
\{(x^n,y^n)\in &\mathcal{X}^n\times\mathcal{Y}^n: d(x^n, \varphi(g(x^n,y^n),y^n))\le\Delta+\delta \}.
\end{align} 
Thus, the inequality in (\ref{tau}) is equivalent to
\begin{align}
Q^n(A)>1-\tau, \label{tau1}
\end{align}
where $Q^n(A)$ is the probability that an i.i.d. randomly generated $(X^n,Y^n)$ according to the distribution $Q$ falls in the set $A$, i.e.,
\begin{align}
Q^n(A)\triangleq \sum_{(x^n,y^n)\in A}\prod_{i=1}^n Q(x_i,y_i).
\end{align}
We further define $A(Q)$ as the intersection of the set $A$ with the typical set $\mathcal{T}_{[Q]_\delta}^n$, where the definition of a typical set can be found in \cite[Definition 2.8]{Csiszar:2011}, i.e.,
\begin{align}
A(Q)\triangleq A\cap\mathcal{T}_{[Q]_\delta}^n.
\end{align}
From \cite[Lemma 2.12]{Csiszar:2011}, we have
\begin{align}
Q^n\left(\mathcal{T}_{[Q]_\delta}^n\right)\ge1-\epsilon_n, \label{epsilon}
\end{align}
and as a result, from (\ref{tau1}) and (\ref{epsilon}), we have
\begin{align}
Q^n(A(Q))\ge Q^n(A)+Q^n\left(\mathcal{T}_{[Q]_\delta}^n\right)-1\ge1-\epsilon_n-\tau.
\end{align}
Thus, from \cite[Lemma 2.14]{Csiszar:2011}, we have
\begin{align}
\frac{1}{n}\log|A(Q)|\ge H(Q)-\epsilon_n.\label{nonsub}
\end{align}
We further define $A_i(Q)\subset A(Q)$ for $i=1,2,\dots,||g||$ as the set of $(x^n,y^n)$ sequences that   satisfy the distortion constraint when mapped to index $i$ at the encoder, i.e.,
\begin{align}
A_i(Q)\triangleq
\{(x^n,y^n)\in\mathcal{T}_{[Q]_\delta}^n:d(x^n, \varphi(i,y^n))\le\Delta+\delta\}. \label{Nan10}
\end{align}
Since all $(x^n, y^n)$ sequence pairs that satisfy the distortion constraint $\Delta+\delta$ under $(g,\varphi)$ has to be mapped to an index $g(x^n,y^n)$, we have
\begin{align}
A(Q)&=\bigcup_{i=1}^{||g||}A_{i}(Q),\label{specii}\\
A_{i}(Q)\cap A_{j}(Q)&=\emptyset\qquad \text{for }i\ne j.
\end{align}
We define $i_o$ as the index of the set $A_{i}(Q)$ with the largest cardinality, i.e.,
\begin{align}
i^o\triangleq\arg\max_{i\in\{1,2,\dots,||g||\}}|A_{i}(Q)|.
\end{align}
Then, we have
\begin{align}
|A_{i^o}(Q)|\ge\frac{|A(Q)|}{||g||}.\label{cov}
\end{align}
Therefore, we have
\begin{align}
-\frac{1}{n}\log P^n(A_{i^o}(Q))
\le& -\frac{1}{n}\log\left(|A_{i^o}(Q)|\min_{(x^n,y^n)\in A_{i^o}(Q)}P^n((x^n,y^n))\right)\nonumber\\
\le&-\frac{1}{n}\log\left(\frac{|A(Q)|\min_{(x^n,y^n)\in A_{i^o}(Q)}P^n((x^n,y^n))}{||g||}\right)\label{Nan02}\\
 \le& -H(Q)+\epsilon_n+R_{SI}(Q,\Delta)+\delta -\frac{1}{n}\log\left(\min_{(x^n,y^n)\in A_{i^o}(Q)}P^n((x^n,y^n))\right)\label{Nan03}\\
\le& -H(Q)+\epsilon_n+R_{SI}(Q,\Delta)+\delta+D(Q||P)+H(Q)+\epsilon_n \label{Nan04}\\
=& D(Q||P)+R_{SI}(Q,\Delta)+\delta+2\epsilon_n, \label{Nan07}
\end{align}
where (\ref{Nan02}) follows from (\ref{cov}), (\ref{Nan03}) follows from  (\ref{WZr}) and (\ref{nonsub}), and 
(\ref{Nan04}) follows from \cite[Lemma 2.6]{Csiszar:2011}, i.e.,~for any $(x^n,y^n) \in \mathcal{T}_{[Q]_\delta}^n$,
\begin{align}
-\frac{1}{n}\log P^n\left((x^n,y^n)\right)\le D(Q||P)-&H(Q)+\epsilon_n.
\label{misseq}
\end{align}
Now, we  construct the deception function $f$ according to $(g,\varphi)$ described above, i.e., 
\begin{align}
f(y^n)=\varphi(i^o,y^n). \label{Nan06}
\end{align}
Then we have
\begin{align}
-\frac{1}{n}\log\mathsf{Pr}(d(X^n,f(Y^n))\le \Delta+\delta) 
=&-\frac{1}{n}\log P^n \left(\{(x^n,y^n): d(x^n,\varphi(i^o,y^n))\le \Delta+\delta\}\right) \label{Nan05}\\ 
=& -\frac{1}{n}\log P^n(A_{i^o}(Q)) \label{Nan08}\\
\leq& D(Q||P)+R_{SI}(Q,\Delta)+\delta+2\epsilon_n, \label{Nan09}
\end{align}
where (\ref{Nan05}) follows from the construction of $f$ in (\ref{Nan06}), (\ref{Nan08}) follows from the definition of $A_i(Q)$ in (\ref{Nan10}), and (\ref{Nan09}) follows from  (\ref{Nan07}). 

Thus, we have shown that there exists a deception function $f$, constructed according to the encoding and decoding function of the corresponding rate distortion problem with side information, that can achieve the the deception exponent $D(Q||P)+R_{SI}(Q,\Delta)$ for any distribution $Q$ defined on $\mathcal{X}\times\mathcal{Y}$ and distortion constraint $\Delta$.
This concludes the proof of the achievability.
\section{The Converse}
In this section, we will prove that for any deception function $f$, the deception exponent can not be smaller than $\min_Q\{D(Q||P)+R_{SI}(Q,\Delta)\}$. This will be proven by contradiction, i.e., we will show that if there is a deception function with the deception exponent equal to $\min_Q\{D(Q||P)+R_{SI}(Q,\Delta)\}-\alpha$ for some $\alpha>0$, then we can construct a coding scheme in the rate distortion with side information problem with achievable rate smaller than $R_{SI}(Q,\Delta)$, which is obviously false. In contrast to  the achievability, which selects one deception function $f(y^n)$ from many rate distortion decoding functions $\phi(i, y^n)$, $i=1,2,\cdots, \|g\|$, i.e., (\ref{Nan06}), the converse requires us to construct many rate distortion decoding functions $\phi(i, y^n)$, $i=1,2,\cdot, \|g\|$ from one deception function $f(y^n)$, which is more difficult. 

We first assume a deception function $f$, which achieves the deception exponent $E$ under distortion constraint $\Delta$, as defined in (\ref{defi}). The proof of the converse includes the following three steps.

\noindent\textbf{Step 1 Type selection:} In this step, we will select one type among all the types, which contributes most to the deception probability for the function $f$.

We define a set $A\subset\mathcal{X}^n\times\mathcal{Y}^n$ as follows
\begin{align}
A\triangleq\left\{(x^n,y^n)\in\mathcal{X}^n\times\mathcal{Y}^n:d(x^n, f(y^n))\le\Delta+\delta\right\}.\label{defA}
\end{align}
Based on the definition of set $A$ and the definition of deception exponent in (\ref{defi}), we have that deception exponent $E$ is achievable is equivalent to
\begin{align}
-\frac{1}{n}\log P^n(A)\le E+\delta. \label{Nan13}
\end{align}
We further define  $\mathcal{P}$ as the set of all possible empirical distribution on $\mathcal{X}^n\times\mathcal{Y}^n$, i.e.,
\begin{align}
\mathcal{P}\triangleq\left\{Q:Q(x,y)=\frac{i}{n}, \text{ for all } (x,y)\in\mathcal{X}\times\mathcal{Y}, i\in\{0,1,2,\dots,n\}\right\}.
\end{align}
For any distribution $Q\in\mathcal{P}$, we define the set $A(Q)$ as the intersection between the set $A$ and $\mathcal{T}_{Q}^n$, where $\mathcal{T}_{Q}^n$ is the type of $Q$ as defined in \cite[Definition 2.1]{Csiszar:2011}, i.e.,
\begin{align}
A(Q)=A\cap\mathcal{T}_{Q}^n.\label{defAQ}
\end{align}
Thus, we have 
\begin{align}
A = \bigcup_{Q\in\mathcal{P}} A(Q).
\end{align}
Since the type $\mathcal{T}_Q^n$ for different empirical distributions $Q\in\mathcal{P}$ are disjoint, we have
\begin{align}
P^n(A)=\sum_{Q\in\mathcal{P}} P^n(A(Q)). \label{Nan11}
\end{align}
We define the type $Q^o$ as the type which contributes most to the deception probability, i.e.,
\begin{align}
Q^o\triangleq \arg\max_{Q\in\mathcal{P}} P^n(A(Q)). \label{Nan11.3}
\end{align}
Let $\beta$ by the number of different types in $\mathcal{X}^n\times\mathcal{Y}^n$. From (\ref{Nan11}) and (\ref{Nan11.3}), we have
\begin{align}
P^n(A(Q^o)) \geq \frac{P^n(A)}{\beta}.\label{Nan11.2}
\end{align}
Thus, we have
\begin{align}
-\frac{1}{n}\log P^n(A(Q^o)) &\leq -\frac{1}{n}\log \frac{P^n(A)}{\beta} \label{Nan11.1}\\
& \leq E+\delta+\frac{1}{n} \log \beta \label{Nan12}\\
&<E+\epsilon_n+\delta, \label{Nan14}
\end{align}
where (\ref{Nan11.1}) follows from (\ref{Nan11.2}), (\ref{Nan12}) follows from (\ref{Nan13}), and (\ref{Nan14}) follows from the fact that $\beta \leq (n+1)^{|\mathcal{X}||\mathcal{Y}|}$ \cite[Lemma 2.2]{Csiszar:2011}.

Since every sequence in the same type is equally probable, we have
\begin{align}
P^n(A(Q^o))=|A(Q^o)| P^n(x^n,y^n), \qquad \forall (x^n,y^n) \in \mathcal{T}_{Q^o}^n.
\end{align}
Thus, for any sequence pair $(x^n,y^n)\in\mathcal{T}_{Q^o}^n$, we have
\begin{align}
\frac{1}{n}\log|A(Q^o)|&=\frac{1}{n} \log \frac{P^n(A(Q^o)}{P^n(x^n,y^n)} \nonumber\\
& > -E-\epsilon_n-\delta-\frac{1}{n}\log P^n((x^n,y^n)) \label{Nan15}\\
&=D(Q^o||P)+H(Q^o)-E-\epsilon_n-\delta,\label{docsize}
\end{align}
where (\ref{Nan15}) follows from (\ref{Nan14}), and (\ref{docsize}) follows from \cite[Lemma 2.6]{Csiszar:2011}, i.e., for distribution $Q^o$,
\begin{align}
-\frac{1}{n}\log P^n((x^n,y^n))=D(Q^o||P)+H(Q^o), \quad \forall (x^n,y^n)\in\mathcal{T}_{Q^o}^n. \nonumber
\end{align}

\textbf{Step 2 Permutation}: In this step, we will construct a rate distortion code, restricted to the type $Q^o$, with side information at both the encoder and decoder from the deception function $f$  via permutations. 

We consider the symmetric group $\mathcal{S}_n$, which consists of all the permutations on $\{1,2,\dots,n\}$. For a set $S\subset\mathcal{X}^n$, and a permutation $\pi\in\mathcal{S}_n$, define the set $\pi(S)$ as the set of sequences that is permuted from the sequences in set $S$ by the permutation $\pi$, i.e., 
\begin{align}
\pi(S)\triangleq\{\bar{x}^n\in\mathcal{X}^n:\exists x^n\in S, \pi(x^n)=\bar{x}^n\}.\label{permS}
\end{align}
Thus, $\pi(S)$ is a permuted version of the set $S$. 

We will use the following lemma, proved by Ahlswede in 1980, to obtain a covering of $\mathcal{T}_{Q^o}^n$ by  the permuted versions of $A(Q^o)$. 
\begin{Lem}[Covering Lemma]  \cite[Section 6.1]{Ahlswede:1980}\label{covlemma}
For any set $S\in\mathcal{T}_{Q}^n$, there exist permutations $\pi_1,\pi_2\dots,\pi_k\in\mathcal{S}_n$ with 
\begin{align}
\bigcup_{i=1}^k\pi_i(S)=\mathcal{T}_{Q}^n,
\end{align}
if
\begin{align}
k>\frac{|\mathcal{T}_{Q}^n|}{|S|}\log|\mathcal{T}_{Q}^n|.
\end{align}
\end{Lem}
From the above lemma, by letting $Q$  be $Q^o$ and $S$  be $A(Q^o)$, we have that there exist permutations $\pi_1,\pi_2\dots,\pi_k\in\mathcal{S}_n$ such that
\begin{align}
\bigcup_{i=1}^k\pi_i(A(Q^o))=\mathcal{T}_{Q^o}^n,\label{typecov}
\end{align}
where $k$ satisfies
\begin{align}
\frac{1}{n}\log k
&=\frac{1}{n} \log |\mathcal{T}_{Q^o}^n|-\frac{1}{n} \log |A(Q^o)|+\log \left(\log |\mathcal{T}_{Q^o}^n| \right)+\epsilon_n \nonumber\\
& \leq E-D(Q^o||P)+2\epsilon_n+\delta+\frac{\log (n H(Q^o))}{n}, \label{Nan16}
\end{align}
where (\ref{Nan16}) follows from (\ref{docsize}) and the fact that
the size of the type $|\mathcal{T}_{Q^o}^n|$ satisfies \cite[Lemma 2.3]{Csiszar:2011}
\begin{align}
|\mathcal{T}_{Q^o}^n|\le\exp(nH(Q^o)).\nonumber
\end{align}
%
%
Based on (\ref{defA}) and (\ref{defAQ}), we have that 
\begin{align}
A(Q^o)=\left\{(x^n,y^n)\in\mathcal{T}^n_{Q^o}:d(x^n, f(y^n))\le\Delta+\delta\right\}
\end{align}
Based on the definition in  (\ref{seqdis}), we see that the same permutation of the two sequences does not change the distortion between the two sequences. Therefore we have for $i=1,2,\dots,k$
\begin{align}
A(Q^o)=\left\{(x^n,y^n)\in\mathcal{T}^n_{Q^o}:d(\pi_i(x^n), \pi_i(f(y^n)))\le\Delta+\delta\right\}.
\end{align}
From the definition of the permutation of a set in (\ref{permS}), we have
\begin{align}
\pi_i(A(Q^o))=\left\{(\bar{x}^n,\bar{y}^n)\in\mathcal{T}^n_{Q^o}:\exists (x^n,y^n)\in A(Q^o),(\pi_i(x^n),\pi_i(y^n))=(\bar{x}^n,\bar{y}^n)\right\}.
\end{align}
Thus, by combining the above two equations, we can view the set $\pi_i(A(Q^o))$ as
\begin{align}
\pi_i(A(Q^o))
&=\left\{(\bar{x}^n,\bar{y}^n)\in\mathcal{T}_{Q^o}^n:d(\bar{x}^n, \pi_i(f(\pi_i^{-1}(\bar{y}^n))))\le\Delta+\delta\right\}.
\end{align}
In other words, the permuted set $\pi_i(A(Q^o))$ can be characterized by a composite function $\pi_i(f(\pi_i^{-1}(\cdot)))$.

With the sets $\pi_i(A(Q^o))$ and the functions $\pi_i(f(\pi_i^{-1}(\cdot)))$ for $i=1,2,\dots,k$, we are ready to construct a rate distortion code with side information. 
 We assume that $(X^n, Y^n)$ are generated i.i.d. according to distribution $Q^o$. We will construct an encoding-decoding  function pair $(g',\varphi')$ for all $(x^n, y^n) \in \mathcal{T}_{Q^o}^n$ as follows. 
 
We define $g'(x^n,y^n)= i$ if $(x^n,y^n)\in\pi_i(A(Q^o))$. If there exist multiple sets $\pi_i(A(Q^o))$ to which $(x^n,y^n)$ belongs, we can arbitrarily pick one set and assign the index of the set to the output of the function $g'$. Define $Q^o_Y$ as the marginal distribution of $Q^o$ in $\mathcal{Y}$. Then, for $i=1,2,\dots,k$ and $y^n\in\mathcal{T}_{Q^o_Y}^n$, we define the decoding function $\varphi'$ as follows
\begin{align}
\varphi'(i,y^n)=\pi_i(f(\pi_i^{-1}(y^n))),\qquad i=1,2,\dots,k.
\end{align}
Due to the covering in (\ref{typecov}), we obtain an encoding-decoding function pair $(g',\varphi')$ for the rate-distortion problem with side information available at both the encoder and decoder for every $(x^n,y^n)\in\mathcal{T}_{Q^o}^n$, which satisfies
\begin{align}
||g'||&=k,\\
d(x^n,\varphi'(g'(x^n,y^n),y^n))&\le \Delta+\delta.
\end{align}

\textbf{Step 3 Blowing-up}: In the previous step, we have construct a code $(g',\varphi')$ for every sequence pair $(x^n,y^n)$ in the type $\mathcal{T}_{Q^o}^n$. In this step, we will expand the code, first to the neighborhood of the type $\mathcal{T}_{Q^o}^n$, and then to the whole space $\mathcal{X}^n\times\mathcal{Y}^n$. The expansion uses the Blowing-up lemma \cite[Chapter 5]{Csiszar:2011}.

First, we expand the code we described in the previous step to the neighborhood of the type $\mathcal{T}_{Q^o}^n$. To do so, let us introduce the definition of the neighborhood of a set as follows.
\begin{Def}\cite[Chapter 5]{Csiszar:2011}
Given a set $S\subset\mathcal{X}^n$, we define the Hamming $l$ neighborhood of $S$ as
\begin{align}
\mathbf{\Gamma}^{l}(S)\triangleq&\left\{x^n\in\mathcal{X}^n:\exists\bar{x}^n\in S, \text{ s.t. }
d_H(x^n,\bar{x}^n)\le l\right\},
\end{align}
where $d_H$ represents the Hamming distance between two sequence pairs, i.e., the number of positions in which the two sequence pairs differ.
\end{Def}

Assume a sequence of positive integer $l_n$ with 
$
\frac{l_n}{n}$ converging to $0$.
We consider the $l_n$ neighborhood of the type $\mathcal{T}_{Q^o}^n$, i.e., $\mathbf{\Gamma}^{l_n}(\mathcal{T}_{Q^o}^n)$. Based on Definition 2 and (\ref{typecov}), we have
\begin{align}
\mathbf{\Gamma}^{l_n}(\mathcal{T}_{Q^o}^n)=\bigcup_{i=1}^k\mathbf{\Gamma}^{l_n}(\pi_i(A(Q^o))). \label{NanLiu01}
\end{align}

For any $(x^n,y^n) \in \mathbf{\Gamma}^{l_n}(\mathcal{T}_{Q^o}^n)$, we will construct a rate distortion code with side information at both the encoder and decoder $(g,\varphi)$ restricted to the $l_n$ neighborhood of the type $\mathcal{T}_{Q^o}^n$, i.e.,~$\mathbf{\Gamma}^{l_n}(\mathcal{T}_{Q^o}^n)$,  from the rate distortion code $(g',\varphi')$ defined on the type $\mathcal{T}_{Q^o}^n$ as described in the previous step. The basic idea is that for $(x^n,y^n)$ in the neighborhood of $(\bar{x}^n,\bar{y}^n)\in\mathcal{T}_{Q^o}^n$, we adopt the encoder-decoder pair $(g',\varphi')$ for $(\bar{x}^n,\bar{y}^n)$ as the encoder-decoder pair for $(x^n,y^n)$. The details are as follows.  

The encoding function $g$ on the set $\mathbf{\Gamma}^{l_n}(\mathcal{T}_{Q^o}^n)$ includes two parts, i.e., $g=(g_1,g_2)$. The function $g_1$ can be constructed 
 in a similar way as we constructed the function $g$ on the type $\mathcal{T}_{Q^o}^n$ in the previous step. More specifically, for any $(x^n,y^n) \in \mathbf{\Gamma}^{l_n}(\mathcal{T}_{Q^o}^n)$, 
we define $g_1(x^n,y^n)= i$ if $(x^n,y^n)\in\mathbf{\Gamma}^{l_n}(\pi_i(A(Q^o)))$. Based on (\ref{NanLiu01}), we know such $i$ always exists. If there exist multiple sets $\mathbf{\Gamma}^{l_n}(\pi_i(A(Q^o)))$ to which $(x^n,y^n)$ belongs, we can arbitrarily pick one set and assign the index of the set to the output of the function $g_1$. Then we have
\begin{align}
||g_1||=k.
\end{align}

Once we determine that $g_1(x^n,y^n)= i$, we can assert that there exists a sequence pair $(\bar{x}^n,\bar{y}^n)\in\pi_i(A(Q^o))$ such that 
\begin{align}
d_H((x^n, y^n),(\bar{x}^n,\bar{y}^n))\le l_n.
\end{align}
We would like to adopt $\varphi'(i,\bar{y}^n)$ as the decoding function $\varphi(i,y^n)$. However, $\bar{y}^n$ is determined based on sequence pair $(x^n,y^n)$. Therefore, the decoder, with only the knowledge of $y^n$, can not determine $\bar{y}^n$ by itself. To overcome this problem, we need to design $g_2$, i.e., the second part of encoding function,  to inform the decoder of $\bar{y}^n$ as follows.
%

Let us consider $y^n$, the second sequence in the pair $(x^n,y^n)$. We construct a $l_n$ Hamming neighborhood $\mathbf{\Gamma}^{l_n}(y^n)$ around the sequence $y^n$, which is called a Hamming ball, and give every sequence in $\mathbf{\Gamma}^{l_n}(y^n)$ an index. 
 Obviously, the sequence $\bar{y}^n$ can be uniquely determined by the sequence $y^n$ together with the index of $\bar{y}^n$ with respect to $y^n$, say $j\in\{1,2,\dots,\left|\mathbf{\Gamma}^{l_n}(y^n)\right|\}$. 
Then $g_2$, the second part of the encoding function $g$, can be defined as the above index, i.e.,
\begin{align}
g_2(x^n,y^n)=j,
\end{align}
and we have
\begin{align}
||g_2||=\left|\mathbf{\Gamma}^{l_n}(y^n)\right|,
\end{align}
where the size of the Hamming ball satisfies \cite[Lemma 5.1]{Csiszar:2011}
\begin{align}
\frac{1}{n}\log\left|\mathbf{\Gamma}^{l_n}(y^n)\right|\le h\left(\frac{l_n}{n}\right)+\frac{l_n}{n}\log|\mathcal{Y}|,\label{hball}
\end{align}
and $h(\cdot)$ represents the binary entropy function.
Therefore, the size of the function $g$ is
\begin{align}
||g||=||g_1||\cdot||g_2||=k\left|\mathbf{\Gamma}^{l_n}(y^n)\right|. \label{NanLiu02}
\end{align}

Note that for $(x^n,y^n)\in\mathbf{\Gamma}^{l_n}(\mathcal{T}_{Q^o}^n)$, we have $y^n\in\mathbf{\Gamma}^{l_n}(\mathcal{T}_{Q^o_Y}^n)$.
Thus, we expand the domain of the second argument of the function $\varphi$ from the type $\mathcal{T}_{Q^o_Y}^n$ to its $l_n$ neighborhood $\mathbf{\Gamma}^{l_n}(\mathcal{T}_{Q^o_Y}^n)$.  
 
Assume the decoder observes $y^n \in\mathbf{\Gamma}^{l_n}(\mathcal{T}_{Q^o_Y}^n)$, and obtain $g(x^n,y^n)$ from the encoder as follows
\begin{align}
g(x^n,y^n)=(i,j).
\end{align}
From $y^n$ and $j$, we can determine the sequence $\bar{y}^n$ in the type $\mathcal{T}_{Q^o_Y}^n$. We then define $\varphi((i,j),y^n)=\varphi'(i,\bar{y}^n)$ for $i=1,2,\dots,k$,
where the decoding function $\varphi'(i,\bar{y}^n)$ was defined in the previous step.
%

Based on the definition of $\mathbf{\Gamma}^{l_n}(\pi_i(A(Q^o)))$ , for every $(x^n, y^n)\in\mathbf{\Gamma}^{l_n}(\pi_i(A(Q^o)))$,  there exists a $(\bar{x}^n,\bar{y}^n)\in\pi_i(A(Q^o))$ such that 
\begin{align}
d_H((x^n, y^n),(\bar{x}^n,\bar{y}^n))\le l_n,
\end{align}
and from the definition of $\pi_i(A(Q^o))$, we know that 
\begin{align}
d(\bar{x}^n, \varphi'(i,\bar{y}^n))
\le\Delta+\delta. \label{Nan50}
\end{align}
Therefore, we have
\begin{align}
d(x^n,\varphi(g(x^n,y^n),y^n))&=d(x^n,\varphi'(g'(\bar{x}^n,\bar{y}^n),\bar{y}^n))\nonumber\\
&\le d(\bar{x}^n,\varphi'(g'(\bar{x}^n,\bar{y}^n),\bar{y}^n))+d_M\frac{l_n}{n},\label{Nan21}
\end{align}
where (\ref{Nan21}) is because $x^n$ and $\bar{x}^n$ at most differ in $l_n$ positions and 
\begin{align}
d_M\triangleq \max_{(x,\hat{x})\in\mathcal{X}\times\hat{\mathcal{X}}} d(x,\hat{x}).
\end{align}
Hence, with the above definition of $(g,\varphi)$, we have that 
\begin{align}
||g||&=k\left|\mathbf{\Gamma}^{l_n}(y^n)\right|,\\
d(x^n,\varphi(g(x^n,y^n),y^n))&\le \Delta+\delta+d_M\frac{l_n}{n}. \label{Nan51}
\end{align}
where (\ref{Nan51}) follows from (\ref{Nan50}) and (\ref{Nan21}).

Finally, we expand the rate-distortion code $(g, \varphi)$ to all $(x^n,y^n) \in \mathcal{X}^n \times \mathcal{Y}^n$. For $(x^n,y^n)\notin\mathbf{\Gamma}^{l_n}(\mathcal{T}_{Q^o}^n)$, we define $g(x^n,y^n)=0$. 
And we define $\varphi(0,y^n)$ to be an arbitrary sequence in $\hat{\mathcal{X}}^n$.

Therefore, for $(g,\varphi)$ defined on $\mathcal{X}^n\times\mathcal{Y}^n$, we have
\begin{align}
\frac{1}{n}\log||g|| &\leq 
E-D(Q^o||P)+3\epsilon_n+\delta+\frac{\log (nH(Q^o))}{n} +h\left(\frac{l_n}{n}\right)+\frac{l_n}{n}\log|\mathcal{Y}|,\label{Nan20}
\end{align}
where (\ref{Nan20}) follows from (\ref{Nan16}), (\ref{NanLiu02}) and (\ref{hball}).

We use blowing up lemma to calculate the average distortion for the rate distortion code we constructed above. We restate the blowing up lemma as follows.

\begin{Lem}[Blowing up]\cite[Lemma 5.4]{Csiszar:2011}
To any finite set $\mathcal{X}$ and sequence $\epsilon_n\rightarrow 0$, there exist a sequence of positive integers $l_n$ with $\frac{l_n}{n}\rightarrow 0$ and a sequence $\tau_n\rightarrow 1$ such that for any distribution $Q$ defined on $\mathcal{X}$ and every $n,A\subset\mathcal{X}^n$
\begin{align}
Q^n(A)\ge\exp(-n\epsilon_n)
\end{align}
implies
\begin{align}
Q^n(\mathbf{\Gamma}^{l_n}(A))\ge\tau_n.
\end{align}
\end{Lem}

In our case, from \cite[Lemma 2.3]{Csiszar:2011}, we have that for any sequence $\epsilon_n\rightarrow0$ and sufficiently large $n$
\begin{align}
(Q^{o})^{n}\left(\mathcal{T}_{Q^o}^n\right)\ge (n+1)^{-|\mathcal{X}||\mathcal{Y}|}\ge \exp(-n\epsilon_n).
\end{align}

Thus, from Blowing up lemma, we have that there exists a sequence $\eta_n\rightarrow1$ such that
\begin{align}
(Q^{o})^{n}\left(\mathbf{\Gamma}^{l_n}\left(\mathcal{T}_{Q^o}^n\right)\right)\ge\eta_n.\label{blup}
\end{align}

Therefore, for $(g,\varphi)$ defined on $\mathcal{X}^n\times\mathcal{Y}^n$,
\begin{align}
\mathsf{Pr}\left(d(x^n,\varphi(g(x^n,y^n),y^n))\le \Delta+\delta+d_M\frac{l_n}{n}\right)\ge(Q^{o})^{n}\left(\mathbf{\Gamma}^{l_n}\left(\mathcal{T}_{Q^o}^n\right)\right)
\ge\eta_n,\label{probdis}
\end{align}
where (\ref{probdis}) follows from (\ref{Nan21}) and (\ref{blup}).
The inequality in (\ref{probdis}) leads to
\begin{align}
\mathsf{E}(d(X^n,\varphi(g(X^n,Y^n),Y^n))&\le
\Delta+\delta+d_M\frac{l_n}{n}+d_M(1-\eta_n).
\end{align}
Thus, if we have a deception function, which under the distortion constraint $\Delta$ can achieve the deception exponent
\begin{align}
E=\min_Q D(Q||P)+R_{SI}(Q,\Delta)-\alpha,
\end{align}
 for some $\alpha>0$,
then we can construct a rate distortion code with side information at both the encoder and decoder, where $(X^n, Y^n)$ is generated i.i.d. according to the distribution $Q^o$, that satisfies
\begin{align}
&\frac{1}{n}\log||g||\nonumber\\
 &\leq 
\min_Q \{D(Q||P)+R_{SI}(Q,\Delta)\}-\alpha-D(Q^o||P)+3\epsilon_n+\delta+\frac{\log (nH(Q^o))}{n} +h\left(\frac{l_n}{n}\right)+\frac{l_n}{n}\log|\mathcal{Y}|\nonumber\\
&\leq R_{SI}(Q^o,\Delta)-\alpha+3\epsilon_n+\delta+\frac{\log (nH(Q^o))}{n} +h\left(\frac{l_n}{n}\right)+\frac{l_n}{n}\log|\mathcal{Y}|,
\end{align}
and
\begin{align}
\mathsf{E}(d(X^n,\varphi(g(X^n,Y^n),Y^n))&\le
\Delta+\delta+d_M\frac{l_n}{n}+d_M(1-\eta_n).
\end{align}
Since the rate distortion function $R_{SI}(Q^o,\Delta)$ is a continuous function of $\Delta$, the above result contradicts with the result in the rate distortion problem with side information at both the encoder and decoder. This concludes the proof of the converse. 
\section{Conclusion}
In this paper, we studied the probability of successful deception of an uncompressed biometric authentication system with side information at the adversary. We found the optimal exponent of the deception probability by providing the proofs of both the achievability and the converse. The results are proved by exploiting a connection between the problem of deception with side information and the rate distortion problem with side information at both the encoder and decoder.

\bibliographystyle{unsrt}
\bibliography{refphd}
\end{document}